\documentclass[11pt,epsf,showpacs,amsmath,amssymb]{revtex4}
\usepackage{graphicx}
\usepackage{dcolumn}
\usepackage{bm}
\usepackage{color}
\usepackage{epsf}

\usepackage{amssymb}
\usepackage{amsmath}
\usepackage{hyperref}
\usepackage{appendix}
\begin{document}

\title{Relativistic Temperature Transformation Revisited, One hundred years after Relativity Theory}

\author{M.Khaleghy$^{1,2}$}
\email[]{MKhalegh@uwo.ca}
\author{F.Qassemi$^{1,3}$}
\email[]{FQassemi@uwo.ca}

\affiliation{1-Department of Physics, Shahid Beheshti University,
Tehran, 19834 Iran\\
2-Department of Physics and Astronomy, University of Western Ontario, London, N6A 3K7, Canada\\
3-Department of Applied Mathematics, University of Western Ontario, London, N6A 5B7, Canada}

\begin{abstract}
An attempt has been made to find a consistent and logical form for
relativistic temperature transformation. Other works in this area
have been discussed. Our approach is based on the kinetic theory
of ideal gases.\\
\end{abstract}
 \maketitle
\section*{Introduction}
According to Galilean Transformation (GT), the speed of light is
different in two inertial frames, moving with constant velocity
with respect to each other. In the late 18th and early 19th
century, there was considerable evidence supporting the idea that
the speed of light is an invariant quantity with respect to all
inertial observers. In 1905, Einstein~\cite{einstein} made the
ground breaking discovery that when transforming physical
quantities from one inertial observer's frame to another, the
Lorentz transformations must be used so that the speed of light
remains an invariant quantity. The principle of relativity
modified the Newton's laws of motion and in a sense unified
concepts like space and time. Since then, much effort has been
devoted to the search for the relativistic forms of other laws of
physics. Many authors have tried to modify other fields, such as
thermodynamics, to render it compatible with the relativity theory.\\
Generally, quantities are transformed in any field of physics by
presuming invariance of some basic quantities or covariance of the
form of laws in that discipline under Lorentz transformation. For
example electric and magnetic fields are transformed by assuming
covariance of Maxwell equations. In our efforts to find the
relativistic temperature  transformation, we have conducted a
fairly extensive survey of the existent literature on the subject.
This included books such as "Feynman Lectures on Physics"
~\cite{fey} as well as numerous other papers published in this
area since the birth of relativity by Planck, Einstein, Tolman,
Ott and many others. (See the reference section for citations).


Generally, it can be seen that the results in the current
literature, has been obtained by postulating the covariance of the
first and/or second law(s) of thermodynamics under Lorentz
transformation. It is fair to say that at this moment and to the
best of our knowledge, lack of distinct experimental evidence, has
prevented a consensus to be reached among experts on the subject.
As we have seen the principle of the invariance of the speed of
light has modified the laws of physics. In the light of this
principle we should take a more careful approach when we are about
to make fundamental assumptions. To avoid ambiguity, the authors
of this paper have chosen a method which involves using equations
and concepts that are verified by experiment and theory in the
relativistic domain, such as relativistic form of energy. We have
sought the clearest and simplest set of assumption.\\ In what
follows, we begin with a rather short historical review (for
complete reviews see the references section), and then
we try to derive our transformation from a new point of view.\\
Our notation is as follows: symbols $T$, $P$, $Q$, $S$, $U$ and $W$
are used for temperature, pressure, heat, entropy, internal energy
and work, respectively.

\section{Historical review}
Einstein and Planck~\cite{einstein2,3} were the first who proposed
a transformation for temperature. They applied principle of the
least action to a moving black-body cavity and assumed the first
and the second laws of thermodynamics (\ref{1stlaw}),
(\ref{2ndlaw}) to be covariant:
\begin{equation}
\label{1stlaw} dU=dQ+dW
\end{equation}
\begin{equation}
\label{2ndlaw} dQ=TdS,
\end{equation}
and they argued for the following set of transformations:
\begin{equation}
T=T'/\gamma \hspace{0.5cm}, \hspace{0.5cm}  Q=Q'/\gamma
\end{equation}
\begin{equation}
S=S' \hspace{0.5cm}, \hspace{0.5cm} P=P',
\end{equation}
where primed quantities denote, the ones measured in the proper
frame of reference that is moving with constant velocity
$\overrightarrow{v}$ with respect to a stationary reference frame
and $\gamma$ is:
\begin{equation}
\gamma=\frac{1}{\sqrt{1-\frac{v^2}{c^2}}}.
\end{equation}
Several years later, a new transformation of temperature was derived
by Eddington~\cite{4} and Blanusa~\cite{5} based on the
covariance of the second law of thermodynamics (\ref{2ndlaw}).\\
In 1963, Ott ~\cite{6} provided a critical overview of the
subject. He made the assertion that the transformations, derived
by Eddington and Balnusa~\cite{4,5} are the correct
transformations for the thermodynamic quantities. These are:
\begin{equation}\label{otts} T=\gamma T' \hspace{0.5cm}, \hspace{0.5cm}Q=\gamma Q' \end{equation}
\begin{equation} S=S' \hspace{0.5cm}, \hspace{0.5cm} P=P'. \end{equation}
However, Landsberg~\cite{7,8} argued that the thermodynamic
quantities that are statistical in nature, namely $T, S, U$ should
not be expected to change for an observer who judges the center of
mass to be undergoing a uniform motion. This approach, leads to
the conclusion that some thermodynamic relationships such as the
second law (\ref{2ndlaw}) are not covariant and results in the
following equations:
\begin{equation} \gamma Q=TdS\end{equation}
\begin{equation} Q=Q'/\gamma\end{equation}
\begin{equation} P=P',\end{equation}
The above discussions can be summarized as the following:
\begin{equation} T=\gamma^a T',\end{equation}
where $a=-1,\;0,\; 1$ comes from Planck/Einstein, Landsberg and
Ott views, respectively.\\
The mentioned proposed transformations for temperature have been
criticized by many authors. Here we briefly review some of the critique.  \\
In 1967, \emph{Moller} ~\cite{13} some years after publishing his book on theory 
of relativity, criticized the issue in a rather detailed paper and called Ott's transformation as
the correct formula, and reminisced accepting of Einstein derivation wrongly as correct formula 
for half a century as strange and unique incident in history of physics.\\
\emph{Landsberg} pointed out that Ott's view as expressed in
(\ref{otts}) requires force to be redefined, a requirement which
is difficult for many to accept ~\cite{14}.\\
\emph{Shizhi} stated that the different formulations are
describing the same physical quantities and are therefore
compatible. He believes that the difference between Einstein's and
Planck's opinions from one hand and Ott's from the other, depends
on the selection of the frame of reference~\cite{15} .\\
\emph{The kinetic theory of ideal gases} has been the basis of
some additional work on the subject in which other relationships
have been proposed~\cite{9,10,11,12}.
\section{Derivation}
The authors of this paper take the appropriate definition of the
relativistic temperature to be one that is based on the kinetic
theory of ideal gases. Before deriving our transformation for
temperature let us try to calculate the relativistic equipartition
theorem. According to Boltzmann probability distribution for a
classical (classic vs. quantum) ideal gas in a thermal bath,
average energy is related to reservoir temperature, $T$, as below:
\begin{equation}
\label{<E>} \langle E \rangle =\bar{E} =
\frac{\int_{-\infty}^{\infty}\int_{-\infty}^{\infty} E e^{-\beta
E}\frac{d^{3}\vec{x}d^{3}\vec{p}}{h^{3}}
}{\int_{-\infty}^{\infty}\int_{-\infty}^{\infty} e^{-\beta
E}\frac{d^{3}\vec{x}d^{3}\vec{p}}{h^{3}}},
\end{equation}
where $h$ is a normalization constant with dimensions of action ,
$\beta=1/(kT)$ and $k$ is the Boltzmann constant. In most
thermodynamics textbooks the above relationship is used to
calculate $\langle E \rangle$. By using the non-relativistic
expression for kinetic energy, $p^2/2m$, it is shown that $\langle
E \rangle = \frac{3}{2}kT$, where the upper limit for the integral
over momentum goes to infinity which requires us to use the
relativistic form of energy.\\ Here, we use the relativistic form
of energy-momentum relationship (\ref{relE}) and
(\ref{a1}), (\ref{a2}) and (\ref{a3}) from Appendix A and invoke
the change of variable (\ref{cv}) to evaluate ~{$\langle E \rangle
$} as below:
\begin{equation} \label{relE}E^2=(pc)^2+(mc^2)^2\end{equation}
\begin{equation} \label{cv} p=mc\hspace{0.1cm}sinh(\chi) \end{equation}
\begin{equation}
\label{averelE}\langle E
\rangle=mc^2(\frac{K_{1}(u)}{K_{2}(u)}+\frac{3}{u}),
\end{equation}
where $u=\beta mc^2$ and $K'$s are modified Bessel functions.\\
Now, as a consistency check it is useful to obtain classical
(u $\gg$ 1) limit of the above relation. From (\ref{averelE}) and (\ref{a4}) in appendix we have:
\begin{equation}\label{besselapr1} K_{1}(u)\simeq(\frac{\pi}{2u})^{1/2} e^{-u} [1+\frac{3}{8u}] \end{equation}
\begin{equation}\label{besselapr2} K_{2}(u)\simeq(\frac{\pi}{2u})^{1/2} e^{-u} [1+\frac{15}{8u}],\end{equation}
by using (\ref{averelE}), (\ref{besselapr1}) and
(\ref{besselapr2}) :
\begin{equation}
\label{classE}\langle E \rangle-mc^2=\langle E_k \rangle
=\frac{3}{2}kT,
\end{equation}
$E_{k}$ denotes kinetic energy.\\

\begin{figure}[b]
  \includegraphics{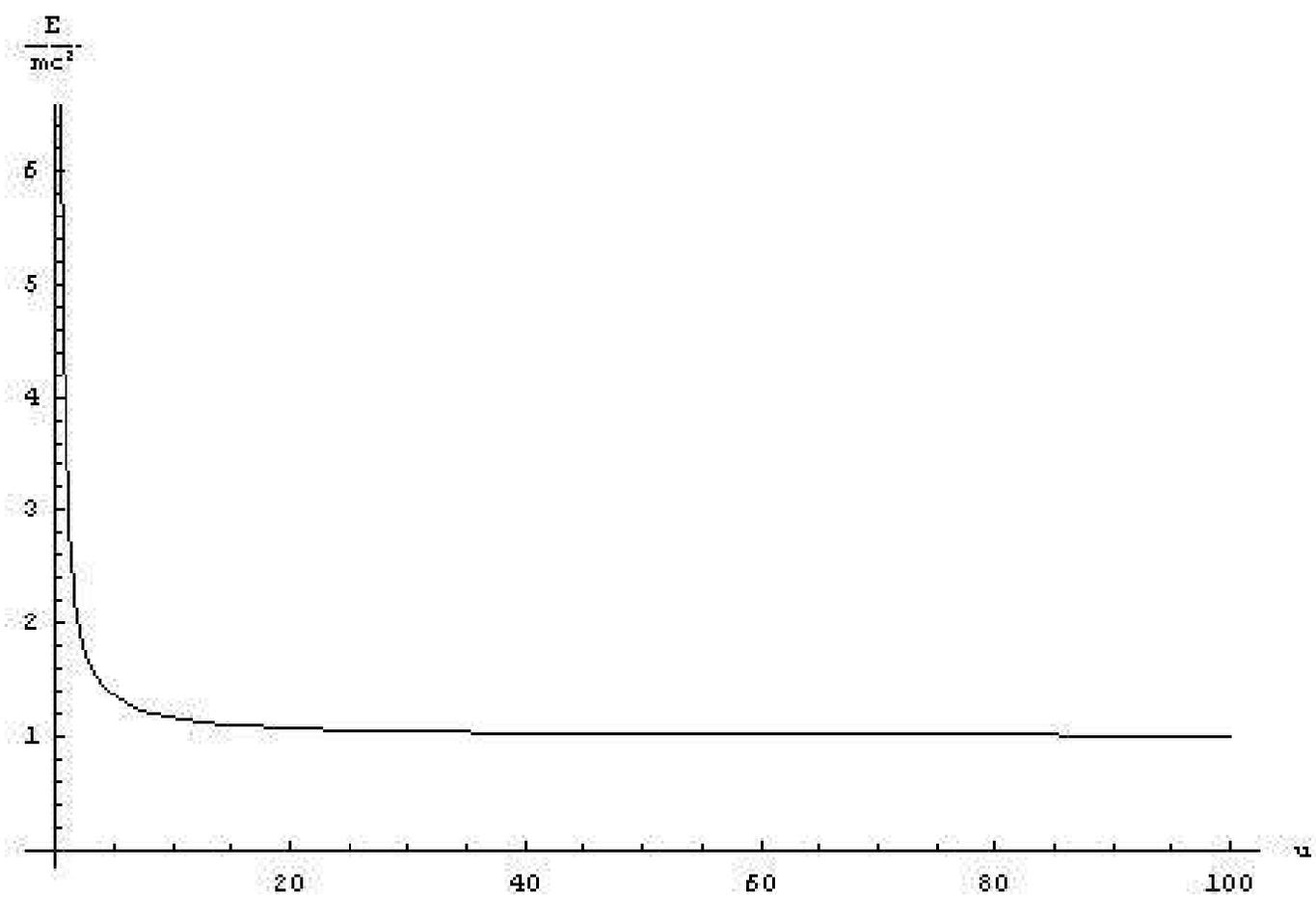}\\
  \caption{Average energy vs $u=\frac{mc^{2}}{kT}$}\label{fig}
\end{figure}

Now, to obtain the relativistic transformation of temperature, an
important question is: What should be kept invariant under Lorentz
transformation?! We presume average energy in any frame of
reference to be related to its temperature by Boltzmann
probability distribution. For clarifying the issue let us explain
our purpose by a useful example. Let the thermodynamic system be,
say, a bottle of helium gas. Strap the bottle onto the seat of a
rocket. As measured in the rocket frame, the temperature of the
helium inside the bottle is $T'$. Now, Let the Rocket be moving
with velocity $\overrightarrow{v}$ relative to the ground-based
lab frame. What is the helium temperature $T$ as measured in the
lab frame compare to $T'$? From 4-vector momentum
transformation for each particle in the bottle we have:
\begin{equation}
\label{relET}E=\gamma (E'+\overrightarrow{v}.\overrightarrow{p'}),
\end{equation}
and for the whole bottle of helium gas, we have: (index $b$
refers to the bottle)
\begin{equation}
\label{relaveET}\langle E \rangle_{b}=\gamma \left(\;\langle E'
\rangle_{b}+\overrightarrow{v}.\langle \overrightarrow{p'} \rangle
\;\right),
\end{equation}
where $\langle \overrightarrow{p'} \rangle $=0 in the rocket frame.\\
Now, substitute average energies by their related temperatures:
\begin{equation}
\label{Ttrans}\frac{K_{1}(u)}{K_{2}(u)}+\frac{3}{u}=\gamma(\frac{K_{1}(u')}{K_{2}(u')}+\frac{3}{u'}),
\end{equation}
where $u=\frac{mc^2}{kT}$ and $u'=\frac{mc^2}{kT'}$. The relation
(\ref{averelE}) as can be seen in the FIG. (\ref{fig}), is a
continuous and monotonically decreasing function. The fact that it
is a single valued function, implies, in eq. (\ref{Ttrans}),
existence of one and just one $T$ corresponding to
$T'$ where $T \geq T'$ because $\gamma\geq 1$.\\
In both classical (u $\gg$ 1) and relativistic (u $\ll$ 1) order
above equation, (\ref{Ttrans}), reduces to Ott transformation
(\ref{otts}).
Asymptotic aspects of the above relation are:\\
\begin{equation}
\gamma \rightarrow 1 \Rightarrow T=T'
\end{equation}
\begin{equation}
\gamma \rightarrow \infty \Rightarrow T=\infty.
\end{equation}
\section{Conclusion}
The transformation derived here using the kinetic theory, applies
to an ideal gas. This result is true for all potentials which
depend only on position (see eq. (\ref{<E>})), and predicts that
moving objects appear hotter to stationary observers. This is in
agreement with Ott's view.\\

At the present time no temperature transformation has been agreed
upon. To reach consensus, it seems necessary that firm
experimental evidence is obtained.\\

\section{Acknowledgement}
We would like to thank Mr. Ardeshir Eftekharzadeh for useful
discussions and helps in preparing this document.

\appendix
\section{Modified Bessel functions}
In this appendix we give some basic mathematical formulae related
to modified Bessel functions $K_{n}(u)$. The expressions are taken
from ~\cite{arfken}:

\begin{eqnarray}
\label{a1} K_{n}(u)&=&
\frac{\sqrt{\pi}}{\Gamma(n+\frac{1}{2})}(\frac{u}{2})^n
\int_{0}^{\infty} e^{-u\hspace{0.1cm}cosh(\chi)} \sinh^{2n}(\chi)
\hspace{0.1cm}d \chi \\
\label{a2}  uK_{n}'(u)-nK_{n}(u)&=&-uK_{n+1}(u)\\
\label{a3}  uK_{n}'(u)+nK_{n}(u)&=&uK_{n-1}(u)\\
\label{a4}  K_{n}(u)&=&(\frac{\pi}{2u})^{1/2}e^{-u}\left[1+\frac{4n^2-1}{1!8u}+\frac{(4n^2-1)(4n^2-9)}{2!(8u)^2}+\cdots
\right]
\end{eqnarray}


\vspace{10cm}
\section*{figures}

\end{document}